**Supernovae and Single-year Anomalies in the Atmospheric Radiocarbon Record**

**Michael Dee[1], Benjamin Pope[2], Daniel Miles[1], Sturt Manning[3] and Fusa Miyake[4]**

[1]University of Oxford – RLAHA, South Parks Road, Oxford OX1 3QY, United Kingdom.
[2]University of Oxford – Physics, Denys Wilkinson Building, Keble Road, Oxford, OX1 3RH, United Kingdom.
[3]Cornell University – Cornell Tree-Ring Laboratory, Ithaca, New York, USA.
[4]Nagoya University – ISEE, Nagoya, Japan.

## Abstract

**Single-year spikes in radiocarbon production are caused by intense bursts of radiation from space. Supernovae emit both high-energy particle and electromagnetic radiation, but it is the latter that is most likely to strike the atmosphere all at once and cause a surge in radiocarbon production. In the 1990s, it was claimed that the supernova in 1006 CE produced exactly this effect. With the radiocarbon spikes in the years 775 and 994 CE now attributed to extreme solar events, attention has returned to the question of whether historical supernovae are indeed detectable using annual radiocarbon measurements. Here, we combine new and existing measurements over six documented and putative supernovae, and conclude that no such astrophysical event has yet left a distinct imprint on the past atmospheric radiocarbon record.**

## Introduction

The rate of natural radiocarbon ($^{14}C$) production is primarily dictated by the abundance of thermalized neutrons in the atmosphere. Their concentration is at its highest in the stratosphere, where they are a secondary product of the incessant cosmic ray (particle) bombardment (see Lal and Peters 1967; Burr 2013). Neutrons of appropriate energy may also be liberated by photonuclear reactions, the most prominent of these effects being the giant dipole resonance (Baldwin & Klaiber 1947; Povinec & Tokar 1970; Pavlov *et al.* 2013), which involves electromagnetic radiation inducing the collective oscillation of all protons against all neutrons in the nucleus. Neutron yields from this effect reach a maximum from photons in the γ-ray region, around 25 MeV (Povinec & Tokar 1970; Pavlov *et al.* 2013). Indeed, it has recently been conjectured that terrestrial gamma ray flashes (TGF) make a minor contribution to atmospheric neutron yields in this fashion (Carlson *et al.* 2010). Radiocarbon is formed by the capture of such neutrons by nitrogen [$^{14}N(n, p)^{14}C$]; other mechanisms are known [such as $^{16}O(n,^{3}He)^{14}C$], but their impact is negligible in comparison (Ligenfelter 1963; Masarik & Beer 2009).

Another potential source of high-energy radiation comes from near-Earth (or galactic) supernovae (SNe). The charged particles emitted by SNe, however, are subject to perturbation by magnetic fields en route to Earth and thus become significantly dispersed and retarded (Melott *et al.* 2015; Güttler *et al.* 2015). In contrast, the γ-ray flux is not impeded in this fashion and arrives in unison with the visible light, which would have appeared as a new star to pre-modern observers. Many types of supernovae exist and their luminosities vary widely - commonly between $10^{46}$ – $10^{49}$ erg (Povinec & Tokar 1970; Miyake *et al*. 2012; Melott *et al.* 2015; Güttler *et al.* 2015). A further complication is that SNe may emit γ-rays isotropically or in a highly collimated fashion, making estimation of their impact on Earth even more difficult.

Damon *et al.* (1995) claimed that a rise in atmospheric $^{14}C$ levels around 1006 CE was attributable to the well-attested Type 1a supernova at this time, denoted SN1006 (supernova in the year 1006 CE). Their study comprised 75 conventional radiocarbon measurements on annual tree-rings between 1000 – 1010 CE. The observed rise in $^{14}C$ (~ 6 ‰) actually peaked some 2 – 3 years after the star was first documented (see Table 1). Whilst this offset was perplexing to the authors, it concurs well with recent modelling of $^{14}C$ transport through the stratosphere and troposphere (Pavlov *et al.* 2013; Levin *et al.* 2010; Güttler *et al.* 2015). Only one attempt has since been made to replicate these findings, and it could not discern any significant uplift around 1006 CE (see Menjo *et al.* 2005). The study also failed to detect SN1054, the explosion that generated the Crab Nebula. Indeed, the authors doubted whether any historical SNe was energetic enough to be visible in the $^{14}C$ record, especially given the ebbs and flows of the Schwabe cycle (Menjo *et al.* 2005).

Attention recently returned to this issue after Miyake *et al.* (2012) reported a rapid increase in atmospheric $^{14}C$ levels in Japanese tree-rings between 774 – 775 CE. The single-year anomaly was of unprecedented magnitude (~ 12 ‰). Just one year later, the same team reported very similar data for the years 993 – 994 CE (Miyake *et al.* 2013). Importantly, the uplifts were only apparent when annual sequences of tree-rings were measured, as opposed to the more common practice of analysing decadal blocks (see Figure 1). Furthermore, it has since been established that the anomalies were globally synchronous and approximately uniform in magnitude. The 775 CE spike has already been uncovered in dendrochronological archives from Germany (Usoskin *et al.* 2013), the USA and Russia (Jull *et al.* 2014), and New Zealand (Güttler *et al.* 2015). Henceforth, these single-year spikes in $^{14}C$ concentration will be referred to as *Miyake Events*.

In addition to their unprecedented abruptness and scale, Miyake Events are also unique because they represent significant increases in $^{14}C$. A myriad of geological and oceanographic processes can drive depletions, but no terrestrial process – prior to the nuclear age – could be responsible for such sharp enrichments. On this basis, as well as their global impact, it was deduced that the spikes must have been the result of intense pulses of radiation from space. At first, the sun was not considered a likely cause, as it was not thought capable of emitting radiation of the required energy, so supernovae and other γ-ray sources were preferred (Miyake *et al.* 2012; Pavlov *et al.* 2013; Hambaryan & Neühauser 2013). However, the consensus now is that intense Solar Energetic Particle (SEP) events were indeed responsible (Melott & Thomas 2012; Thomas *et al.* 2013; Usoskin *et al.* 2013; Güttler *et al.* 2015; Mekhaldi *et al.* 2015). SEPs either arise because of extreme solar flares or Interplanetary Coronal Mass Ejections (ICMEs). A supernova origin has now effectively been discounted, on two main grounds. Firstly, no historical observations exist for supernovae around 775 or 994 CE; although, the expected galactic SN rate of ~ 1 – 2 per century does suggest that many past events have gone undetected (Tammann *et al.* 1994). As is shown in Table 1, only a handful of observations do exist, and none of them pertain to the night sky of the Southern Hemisphere. Secondly, no Galactic supernova remnant can be attributed to an event at either of these dates.
The aim of this study is to establish categorically whether any historical SNe can be detected in the past atmospheric $^{14}C$ record.

## Methods

We combined new and existing $^{14}C$ measurements on annual tree-rings that traversed the following historical astronomical records.

*1. Star of Bethlehem (SB)*

This short-lived star is mentioned twice in the gospel of Matthew. Its historicity and date have long been debated (Tipler 2005), with recent studies centring on 5 BCE (Kidger 1999). For this project, we measured new single rings of oak (*Quercus robur*) dendrochronologically dated to the years 6 – 1 BCE from the Roman-British archaeological site of Hacheston (Miles pers comm.).

*2. SN185*

The appearance of a *kèxīng* or 'guest star' in 185 CE is recorded in the *Houhanshu* (History of the Later Han Dynasty) of Imperial China. Although commonly referred to as the earliest observation of a supernova, this conclusion is by no means unanimous with some palaeographers suggesting the text describes a comet (Schaefer 1995; Chin & Huang 1994; Strom 2008; Zhao *et al.* 2006; Stephenson 2015). For this event, we measured new single rings of sequoia (*Sequoiadendron giganteum*), dendrochronologically dated to the years 183 – 188 CE, from King's Canyon National Park, USA.

*3. SN1006*

The supernova in 1006 CE was widely recorded in both the Eastern and Western hemispheres (Stephenson *et al.* 1976; Green & Stephenson 2003). It is thought to have been the brightest star ever witnessed on Earth in historical time (Stephenson *et al.* 1977). We measured new single rings of oak (*Quercus robur*), dendrochronologically dated to the years 1004 – 1010 CE, originally cored from beams in Salisbury Cathedral (Miles 2002). These results were combined with previously published data from Damon *et al.* (1995) and Menjo *et al.* (2005).

*4. SN1054*

This stellar explosion in the *Taurus* constellation was observed in China in July 1054 (Green & Stephenson 2003). Its remnant gas clouds now form the Crab Nebula. For this event, we utilise the published results of Menjo *et al.* (2005).

*5. SN1572 (Tycho's Supernova)*

This supernova is named for the Danish astronomer, Tycho Brahe, who witnessed the appearance of the star in $\beta$ Cassiopeiae in early 1572 CE and published his observations the following year. For this event, we utilise the single-year tree-ring data of IntCal13 (Reimer *et al.* 2013), which extend back to the mid-16$^{th}$ century CE.

*6. SN1604 (Kepler's Supernova)*
The last near-Earth SN to be observed on Earth was more than 400 years ago, in 1604 CE. Although extensively documented around the world, the most renowned observations were made by Johannes Kepler in his publication *Stella Nova in Pede Serpentarii* (Kepler 1606). Once more, the single-year tree-ring data of IntCal13 (Reimer *et al.* 2013) are utilised for this event.

The tree-rings obtained for this work by the Oxford Radiocarbon Accelerator Unit (ORAU) for the SB and SN185 were treated to α-cellulose in accordance with recently published protocols (Staff *et al.* 2014). The samples for SN1006 were given the standard pre-treatment for wood samples (Brock *et al.* 2010). All the cellulosic fractions extracted were combusted, graphitised and measured on ORAU's AMS system, as described in Brock *et al.* (2010) and Bronk Ramsey *et al.* (2004).

## Results
The new $\Delta^{14}C$ measurements obtained by ORAU, together with all the previously published data used in this study, are given in Tables S1 and S2 in the supplementary online material. The new and existing data are summarised in Table 2, and graphically in Figure 2, for the 5 years leading up to and 10 years following each historical observation. Weighted averages were produced for the three data sets available for SN1006. In one sense, this is not the most effective means of determining whether an uplift occurred at this time, as the absolute data come from different species, and different parts of the Northern Hemisphere. However, if a spike did occur, it should be synchronous across the hemisphere so yearly averaging would not affect this pattern. Nonetheless, the three data sets available for SN1006 are also given independently in Table S1 and Figure S1 of the supplementary online material.

## Discussion
Whilst the amalgamated data sets presented here do reveal the natural year-on-year undulation in atmospheric $^{14}C$ concentration, the trends exhibited by the $\Delta^{14}C$ traces in Figure 2 stand in stark contrast to the Miyake Event depicted in Figure 1. If anything, a levelling or gradual decrease in atmospheric $^{14}C$ levels can be discerned in the data for the ten years following each historical observation. It is important to emphasise that the observation dates of the supernovae in the second millennium CE are exactly known, although the evidence pertaining to the earlier events is more equivocal. Thus, any rise in $^{14}C$, which *predates* the historical observation, as can be seen in the profile relating to SN1054, cannot be causally linked with the stellar explosion. The gamma rays from the supernova would arrive at the same time as the visible light, and any potential impact on $^{14}C$ levels would only be evident after this point in time.

Despite the lack of any distinct spikes in the data, the precision of individual $^{14}C$ measurements remains an issue. It is possible that the γ-ray flux from these SNe did increase $^{14}C$ production by <1‰, and the resultant shifts are simply not detectable by this approach. Moreover, although improvements to AMS precision are proceeding apace, distinguishing anomalies at such levels of sensitivity is not thought likely in the foreseeable future. Indeed, it is not possible yet to define which precise radiation-producing events may be detectable by this method. As alluded to earlier, the causes of gamma-ray impacts on the Earth are many and varied and their impacts hard to resolve. For example, even if a more pronounced single-year rise is detected in future, it cannot automatically be assumed that a supernova is not the cause. On the contrary, Miyake Events are thought to represent the upper end of solar emissions (Eichler and Mordecai 2012; Usoskin *et al.* 2013; Cliver *et al.* 2014), which implies that upsurges of greater magnitude may require extra-solar explanations. An intense pulse of γ-rays from a very nearby SN, should remain a possible cause, especially when surveying data over kiloyear timescales. Distinguishing evidence may be found using other proxies. For example, it has long been hypothesized that intense bursts of high-energy γ-flux would also be accompanied by ozone depletion, on account of increased initiation of nitrogen radicals in the atmosphere (Ruderman 1974). However, the search for geochemical and palaeoecological evidence in support of these hypotheses has also proven inconclusive, or implied extremely low rates of occurrence (Reid *et al.* 1978; Ellis and Schramm 1995; Benitez *et al.* 2002; Gehrels *et al.* 2003).

With regard to the exact mechanisms behind Miyake Events, however, the approach applied here may provide further important information. It has already been speculated that the 775 CE event, may be more accurately described as a 'superflare'. Using Kepler photometry, Maehara *et al.* (2012) showed that superflares are common on sun-like stars. Determining whether this is true also of the sun, and what might be driving such superflares, is an active topic of research. As noted by Melott & Thomas (2012), if the 775 CE anomaly was caused by a solar superflare, a recurrence may pose a significant threat to modern technological civilisation, potentially destroying satellites and Earth-bound electrical infrastructure. From Kepler analysis of oscillations in stellar superflares (Balona *et al.* 2015) and associated starspot-related photometric variability (Notsu *et al*. 2013; Maehara *et al.* 2015), it appears likely that superflares, like lesser flares, are powered by the energy stored in a star's magnetic field configuration. It is not yet clear, however, if these are occur on the sun as rare events drawn from the same distribution as ordinary solar flares, or if the occurrence of superflares is confined to younger stars (Wichmann *et al.* 2014). A long-term radioisotope record of solar activity, including Miyake Events, will help answer this question.

## Conclusion

In contrast with Damon *et al.* (1995), we have uncovered no evidence that SN1006 or any of five other historical or putative SNe caused detectable uplifts in the atmospheric concentrations of $^{14}C$. However, this approach still retains enormous potential for elucidating the origin and nature of past radiation impacts on Earth.

## Acknowledgements

The $^{14}C$ measurements obtained by ORAU this work were funded by the Balliol Interdisciplinary Institute. M. W. Dee is supported by a Leverhulme Trust Early Career Fellowship.

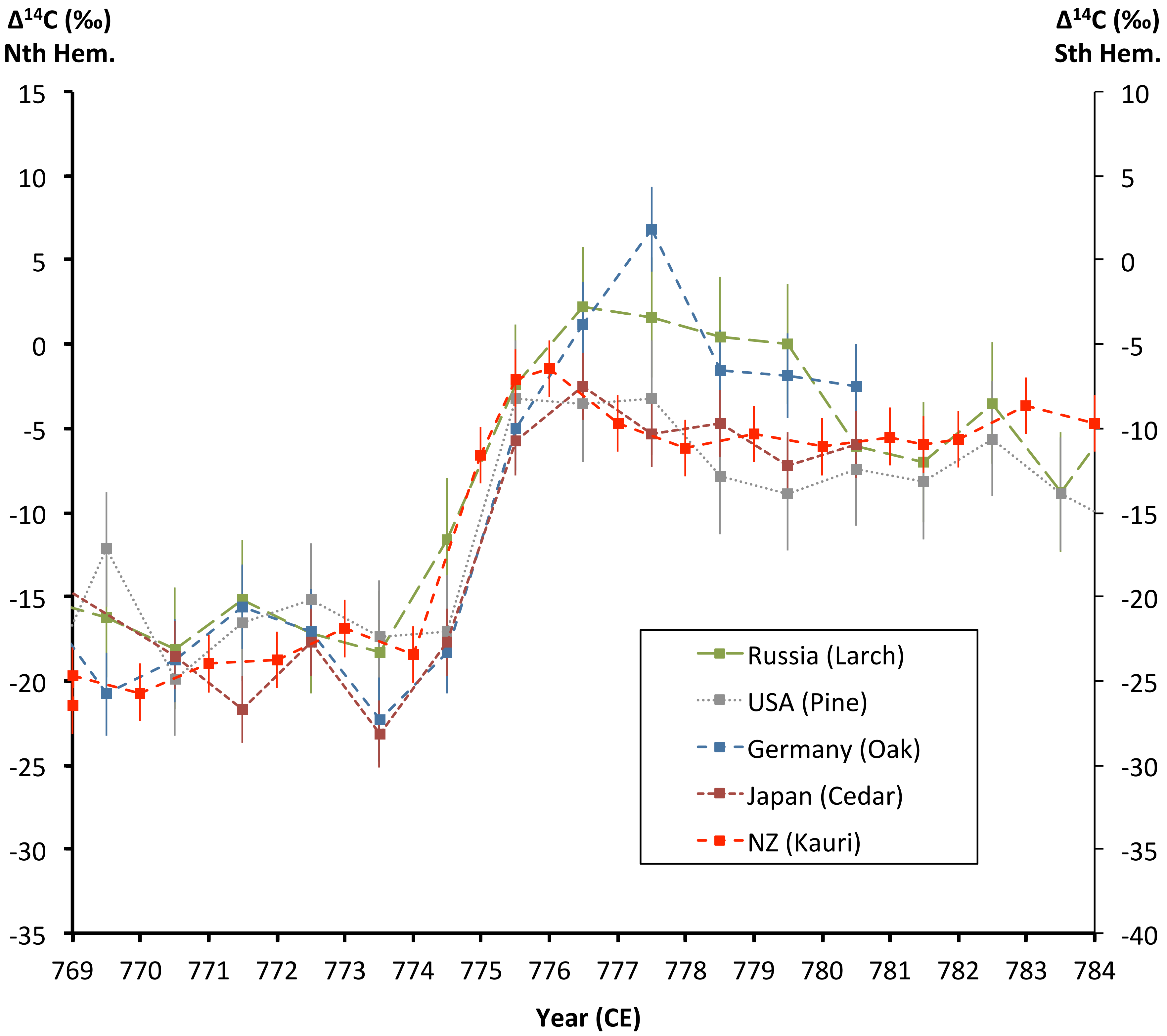

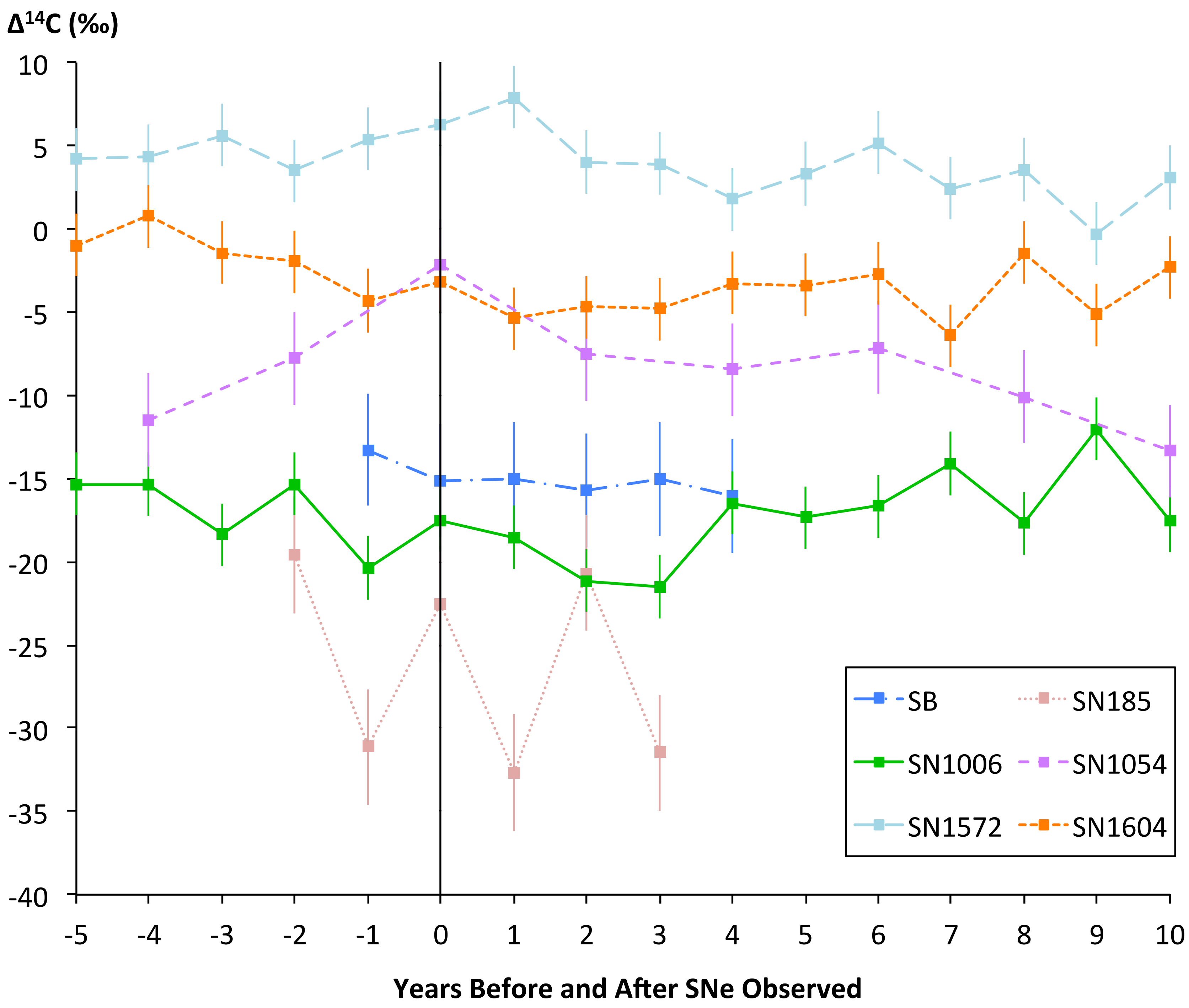

**Figure 1. Published $\Delta^{14}C$ data on the Miyake Event in 775 CE. The 4 Northern Hemisphere data sets [Japan (Miyake *et al.* 2012); Germany (Usoskin *et al.* 2013); USA and Russia (Jull *et al.* 2013) pertain to the left hand axis, and the New Zealand data (Güttler *et al.* 2015) to the right hand axis. The latter is offset by 5 ‰ to account for the differences in absolute activity in the two hemispheres.**

**Figure 2. New and collated $\Delta^{14}C$ data over historical observations of known or potential near-Earth supernovae. The horizontal axis is divided into the 5 calendar years leading up to the observation and the 10 years after it.**

| Name | Date (AD) | Distance (kpc) | Type | Historical Documentation |
|---|---|---|---|---|
| SB | ~ 4 BC | - | - | Biblical tradition |
| SN185 | 185 | 1.8 | - | Observed in China |
| SN386 | 386 | - | - | Observed in China |
| SN393 | 393 | 0.5 | - | Observed in China |
| SN1006 | 1006 | 1.6 | Ia | Observed in Asia & Europe |
| SN1054 | 1054 | 2.0 | II | Widely observed in Asia |
| SN1181 | 1181 | 2.6 | - | Observed in China & Japan |
| SN1572 | 1572 | 2.5 | Ia | Widely observed (studied by Tycho Brahe) |
| SN1604 | 1604 | 1.8 | - | Widely observed (studied by Johannes Kepler) |

**Table 1 Historical records of ephemeral stars thought to be galactic supernovae. The observational records come from Tse-Tsung (1957) and Green and Stephenson (2003); the distances from Earth for SN185 and SN393 come from Damon et al. (1995) and the remainder from Firestone (2014), but estimates vary widely.**

| Name | Year (AD) | Weighted Averages | | Data Incorporated |
|---|---|---|---|---|
| | | $\Delta^{14}C$ (‰) | ± (σ) | |
| SB | 6 BC | -13.3 | 3.6 | This work |
| | 5 BC | -15.1 | 3.8 | |
| | 4 BC | -15.0 | 3.7 | |
| | 3 BC | -15.7 | 3.6 | |
| | 2 BC | -15.0 | 2.4 | |
| | 1 BC | -16.1 | 3.6 | |
| SN185 | 183 | -19.6 | 3.5 | This work |
| | 184 | -31.1 | 3.5 | |
| | 185 | -22.5 | 3.4 | |
| | 186 | -32.7 | 3.4 | |
| | 187 | -20.7 | 3.5 | |
| | 188 | -31.5 | 3.5 | |
| SN1006 | 1001 | -15.3 | 2.5 | This work, Damon *et al.* (1995), and Menjo *et al.* (2005) |
| | 1002 | -15.4 | 2.5 | |
| | 1003 | -18.4 | 2.5 | |
| | 1004 | -15.3 | 1.8 | |
| | 1005 | -20.4 | 1.8 | |
| | 1006 | -17.5 | 1.8 | |
| | 1007 | -18.5 | 1.8 | |
| | 1008 | -21.2 | 1.5 | |
| | 1009 | -21.5 | 1.5 | |
| | 1010 | -16.5 | 1.6 | |
| | 1011 | -17.3 | 1.7 | |
| | 1012 | -16.7 | 1.7 | |
| | 1013 | -14.1 | 1.7 | |
| | 1014 | -17.7 | 2.1 | |
| | 1015 | -12.0 | 2.3 | |
| | 1016 | -17.5 | 2.2 | |
| SN1054 | 1050 | -11.5 | 2.8 | Menjo *et al.* (2005) |
| | 1052 | -7.8 | 2.8 | |
| | 1054 | -2.2 | 2.8 | |
| | 1056 | -7.5 | 2.8 | |
| | 1058 | -8.4 | 2.8 | |
| | 1060 | -7.1 | 2.8 | |
| | 1062 | -10.1 | 2.8 | |
| | 1064 | -13.3 | 2.8 | |
| SN1572 | 1567 | 4.2 | 1.7 | Reimer *et al.* (2013) |
| | 1568 | 4.4 | 2.5 | |
| | 1569 | 5.6 | 2.3 | |
| | 1570 | 3.5 | 1.7 | |
| | 1571 | 5.4 | 2.5 | |
| | 1572 | 6.3 | 2.3 | |
| | 1573 | 7.9 | 2.5 | |
| | 1574 | 4.0 | 1.2 | |
| | 1575 | 3.9 | 2.3 | |
| | 1576 | 1.8 | 2.2 | |
| | 1577 | 3.3 | 2.5 | |
| | 1578 | 5.2 | 2.3 | |
| | 1579 | 2.4 | 2.2 | |
| | 1580 | 3.6 | 2.3 | |
| | 1581 | -0.3 | 2.2 | |
| | 1582 | 3.1 | 2.3 | |
| SN1604 | 1599 | -1.0 | 2.2 | Reimer *et al.* (2013) |
| | 1600 | 0.8 | 1.6 | |
| | 1601 | -1.4 | 1.7 | |
| | 1602 | -2.0 | 2.5 | |
| | 1603 | -4.3 | 2.5 | |
| | 1604 | -3.2 | 1.1 | |
| | 1605 | -5.4 | 1.8 | |
| | 1606 | -4.7 | 1.5 | |
| | 1607 | -4.8 | 2.6 | |
| | 1608 | -3.2 | 1.8 | |
| | 1609 | -3.4 | 1.7 | |
| | 1610 | -2.7 | 2.5 | |
| | 1611 | -6.4 | 1.8 | |
| | 1612 | -1.4 | 2.5 | |
| | 1613 | -5.2 | 2.2 | |
| | 1614 | -2.3 | 1.2 | |

**Table 2 The six astronomical records investigated in this study. Where available, data are given for the 5 years leading up to the first observation and the 10 years thereafter. Weighted averages were calculated for SN1006, as multiple data sets were available for this event. The supplementary online material gives details of all the underlying data (Table S1), as well as the new results expressed as conventional $^{14}C$ ages (Table S2).**

**Figure S1. The three data sets that were combined to form one trace for SN1006**

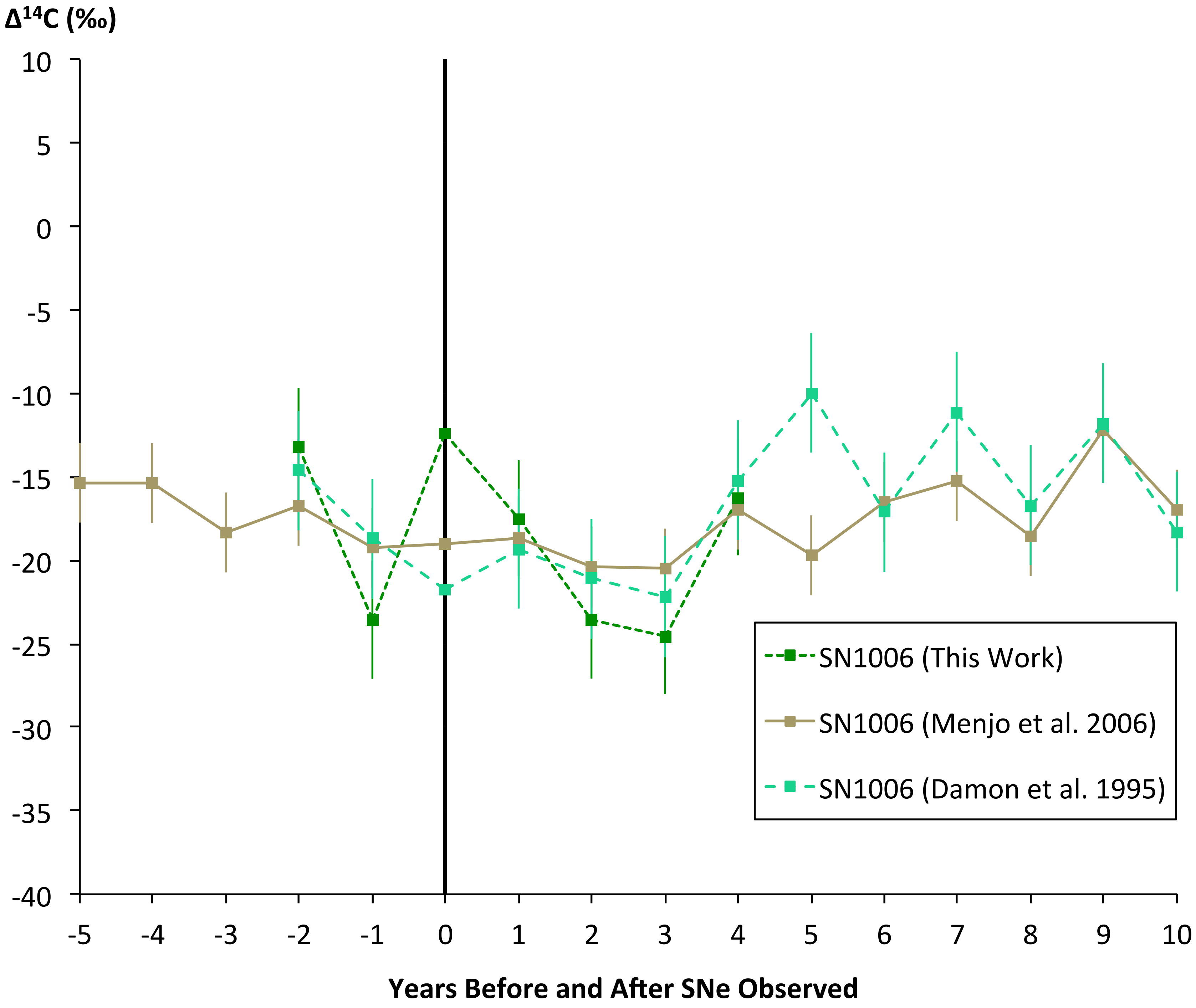

| Table S1: All Data | | | | | | | | | | | | | | | | | | | | | | | | |
|---|---|---|---|---|---|---|---|---|---|---|---|---|---|---|---|---|---|---|---|---|---|---|---|---|
| Year | | ORAU | | Damon *et al.* (1995) | | Menjo *et al.* (2005) | | Miyake *et al.* (2012) | | Reimer *et al.* (2013) | | Jull *et al.* (2013, Russia) | | Jull *et al.* (2013, USA) | | Miyake *et al.* (2013) | | Usoskin *et al.* (2013) | | Miyake *et al.* (2014) | | Weighted Averages | | |
| (AD) | Relative | Δ14C (‰) | ± | Δ14C (‰) | ± | Δ14C (‰) | ± | Δ14C (‰) | ± | Δ14C (‰) | ± | Δ14C (‰) | ± | | | Δ14C (‰) | ± | Δ14C (‰) | ± | Δ14C (‰) | ± | Δ14C (‰) | ± | |
| 6 BC | -1 | -13.3 | 3.6 | | | | | | | | | | | | | | | | | | | **-13.3** | **3.6** | ORAU |
| 5 BC | 0 | -15.1 | 3.8 | | | | | | | | | | | | | | | | | | | **-15.1** | **3.8** | ORAU |
| 4 BC | 1 | -15.0 | 3.7 | | | | | | | | | | | | | | | | | | | **-15.0** | **3.7** | ORAU |
| 3 BC | 2 | -15.7 | 3.6 | | | | | | | | | | | | | | | | | | | **-15.7** | **3.6** | ORAU |
| 2 BC | 3 | -15.0 | 2.4 | | | | | | | | | | | | | | | | | | | **-15.0** | **2.4** | ORAU |
| 1 BC | 4 | -16.1 | 3.6 | | | | | | | | | | | | | | | | | | | **-16.1** | **3.6** | ORAU |
| 183 | -2 | -19.6 | 3.5 | | | | | | | | | | | | | | | | | | | **-19.6** | **3.5** | ORAU |
| 184 | -1 | -31.1 | 3.5 | | | | | | | | | | | | | | | | | | | **-31.1** | **3.5** | ORAU |
| 185 | 0 | -22.5 | 3.4 | | | | | | | | | | | | | | | | | | | **-22.5** | **3.4** | ORAU |
| 186 | 1 | -32.7 | 3.4 | | | | | | | | | | | | | | | | | | | **-32.7** | **3.4** | ORAU |
| 187 | 2 | -20.7 | 3.5 | | | | | | | | | | | | | | | | | | | **-20.7** | **3.5** | ORAU |
| 188 | 3 | -31.5 | 3.5 | | | | | | | | | | | | | | | | | | | **-31.5** | **3.5** | ORAU |
| 769 | -5 | | | | | | | | | | | -16.2 | 4.8 | -12.2 | 3.6 | | | -20.8 | 2.4 | | | **-17.9** | **1.8** | Jull *et al.* 2013; Usoskin *et al.* 2013 |
| 770 | -4 | | | | | | | -18.5 | 1.2 | | | -18.1 | 3.5 | -19.9 | 3.5 | | | -18.8 | 2.2 | | | **-18.6** | **1.0** | Miyake *et al.* 2012; Jull *et al.* 2013; Usoskin *et al.* 2013 |
| 771 | -3 | | | | | | | -21.7 | 2.7 | | | -15.2 | 4.0 | -16.6 | 3.5 | | | -15.6 | 2.2 | | | **-17.4** | **1.4** | Miyake *et al.* 2012; Jull *et al.* 2013; Usoskin *et al.* 2013 |
| 772 | -2 | | | | | | | -17.7 | 1.2 | | | -17.2 | 3.5 | -15.2 | 3.6 | | | -17.1 | 2.7 | | | **-17.4** | **1.0** | Miyake *et al.* 2012; Jull *et al.* 2013; Usoskin *et al.* 2013 |
| 773 | -1 | | | | | | | -23.2 | 2.8 | | | -18.3 | 3.5 | -17.4 | 3.5 | | | -22.3 | 2.5 | | | **-20.9** | **1.5** | Miyake *et al.* 2012; Jull *et al.* 2013; Usoskin *et al.* 2013 |
| 774 | 0 | | | | | | | -17.7 | 1.5 | | | -11.6 | 3.5 | -17.1 | 3.5 | | | -18.3 | 2.8 | | | **-17.1** | **1.2** | Miyake *et al.* 2012; Jull *et al.* 2013; Usoskin *et al.* 2013 |
| 775 | 1 | | | | | | | -5.8 | 1.8 | | | -2.4 | 4.9 | -3.2 | 3.5 | | | -5.0 | 2.8 | | | **-5.0** | **1.3** | Miyake *et al.* 2012; Jull *et al.* 2013; Usoskin *et al.* 2013 |
| 776 | 2 | | | | | | | -2.5 | 1.5 | | | 2.2 | 3.6 | -3.6 | 3.6 | | | 1.2 | 2.7 | | | **-1.4** | **1.2** | Miyake *et al.* 2012; Jull *et al.* 2013; Usoskin *et al.* 2013 |
| 777 | 3 | | | | | | | -5.3 | 1.8 | | | 1.6 | 3.5 | -3.2 | 3.5 | | | 6.8 | 2.8 | | | **-1.5** | **1.3** | Miyake *et al.* 2012; Jull *et al.* 2013; Usoskin *et al.* 2013 |
| 778 | 4 | | | | | | | -4.7 | 1.5 | | | 0.4 | 3.8 | -7.9 | 2.9 | | | -1.6 | 2.8 | | | **-4.2** | **1.1** | Miyake *et al.* 2012; Jull *et al.* 2013; Usoskin *et al.* 2013 |
| 779 | 5 | | | | | | | -7.2 | 1.8 | | | 0.0 | 3.6 | -8.9 | 2.8 | | | -1.9 | 2.8 | | | **-5.6** | **1.2** | Miyake *et al.* 2012; Jull *et al.* 2013; Usoskin *et al.* 2013 |
| 780 | 6 | | | | | | | -6.0 | 1.8 | | | -6.1 | 3.0 | -7.4 | 2.9 | | | -2.5 | 2.8 | | | **-5.6** | **1.2** | Miyake *et al.* 2012; Jull *et al.* 2013; Usoskin *et al.* 2013 |
| 781 | 7 | | | | | | | | | | | -7.0 | 2.7 | -8.2 | 3.1 | | | | | | | **-7.5** | **2.0** | Jull *et al.* 2013 |
| 782 | 8 | | | | | | | | | | | -3.5 | 3.1 | -5.6 | 4.0 | | | | | | | **-4.3** | **2.5** | Jull *et al.* 2013 |
| 783 | 9 | | | | | | | | | | | -8.8 | 3.5 | -8.9 | 2.7 | | | | | | | **-8.9** | **2.1** | Jull *et al.* 2013 |
| 784 | 10 | | | | | | | | | | | -3.3 | 2.9 | -11.0 | 2.8 | | | | | | | **-7.3** | **2.0** | Jull *et al.* 2013 |
| 988 | -5 | | | | | | | | | | | | | | | | | | | -21.9 | 1.8 | **-21.9** | **1.8** | Miyake et al. 2013; Miyake et al. 2014 |
| 989 | -4 | | | | | | | | | | | | | | | -20.2 | 1.9 | | | -21.3 | 1.8 | **-20.8** | **1.3** | Miyake et al. 2013; Miyake et al. 2014 |
| 990 | -3 | | | | | | | | | | | | | | | -25.3 | 2.9 | | | -22.5 | 1.7 | **-23.2** | **1.5** | Miyake et al. 2013; Miyake et al. 2014 |
| 991 | -2 | | | | | | | | | | | | | | | -21.5 | 1.5 | | | -23.1 | 1.9 | **-22.1** | **1.2** | Miyake et al. 2013; Miyake et al. 2014 |
| 992 | -1 | | | | | | | | | | | | | | | -22.8 | 2.0 | | | -24.4 | 1.9 | **-23.6** | **1.4** | Miyake et al. 2013; Miyake et al. 2014 |
| 993 | 0 | | | | | | | | | | | | | | | -20.7 | 1.6 | | | -25.3 | 1.7 | **-22.9** | **1.2** | Miyake et al. 2013; Miyake et al. 2014 |
| 994 | 1 | | | | | | | | | | | | | | | -11.5 | 2.0 | | | -14.0 | 1.8 | **-12.9** | **1.3** | Miyake et al. 2013; Miyake et al. 2014 |
| 995 | 2 | | | | | | | | | | | | | | | -12.9 | 1.5 | | | -18.1 | 1.7 | **-15.2** | **1.1** | Miyake et al. 2013; Miyake et al. 2014 |
| 996 | 3 | | | | | | | | | | | | | | | -11.3 | 2.0 | | | -16.7 | 1.7 | **-14.4** | **1.3** | Miyake et al. 2013; Miyake et al. 2014 |
| 997 | 4 | | | | | | | | | | | | | | | -14.2 | 2.0 | | | -16.4 | 1.7 | **-15.5** | **1.3** | Miyake et al. 2013; Miyake et al. 2014 |
| 998 | 5 | | | | | | | | | | | | | | | -15.3 | 2.0 | | | | | **-15.3** | **2.0** | Miyake et al. 2013; Miyake et al. 2014 |
| 999 | 6 | | | | | | | | | | | | | | | -13.3 | 1.6 | | | | | **-13.3** | **1.6** | Miyake et al. 2013; Miyake et al. 2014 |
| 1001 | 8 | | | | | | | | | | | | | | | -15.0 | 2.9 | | | | | **-15.0** | **2.9** | Miyake et al. 2013; Miyake et al. 2014 |
| 1003 | 10 | | | | | | | | | | | | | | | -17.0 | 2.6 | | | | | **-17.0** | **2.6** | Miyake et al. 2013; Miyake et al. 2014 |
| 1001 | -5 | | | | | -15.3 | 2.5 | | | | | | | | | | | | | | | **-15.3** | **2.5** | Menjo *et al.* 2005 |
| 1002 | -4 | | | | | -15.4 | 2.5 | | | | | | | | | | | | | | | **-15.4** | **2.5** | Menjo *et al.* 2005 |
| 1003 | -3 | | | | | -18.4 | 2.5 | | | | | | | | | | | | | | | **-18.4** | **2.5** | Menjo *et al.* 2005 |
| 1004 | -2 | -13.2 | 3.4 | -14.6 | 3.9 | -16.8 | 2.5 | | | | | | | | | | | | | | | **-15.3** | **1.8** | ORAU; Damon *et al.* 1995; Menjo *et al.* 2005 |
| 1005 | -1 | -23.6 | 3.4 | -18.7 | 4.0 | -19.3 | 2.5 | | | | | | | | | | | | | | | **-20.4** | **1.8** | ORAU; Damon *et al.* 1995; Menjo *et al.* 2006 |
| 1006 | 0 | -12.5 | 3.2 | -21.8 | 4.0 | -19.0 | 2.5 | | | | | | | | | | | | | | | **-17.5** | **1.8** | ORAU; Damon *et al.* 1995; Menjo *et al.* 2007 |
| 1007 | 1 | -17.5 | 3.8 | -19.3 | 3.7 | -18.6 | 2.5 | | | | | | | | | | | | | | | **-18.5** | **1.8** | ORAU; Damon *et al.* 1995; Menjo *et al.* 2008 |
| 1008 | 2 | -23.6 | 3.4 | -21.1 | 3.4 | -20.4 | 1.9 | | | | | | | | | | | | | | | **-21.2** | **1.5** | ORAU; Damon *et al.* 1995; Menjo *et al.* 2009 |
| 1009 | 3 | -24.5 | 3.5 | -22.2 | 3.8 | -20.5 | 1.9 | | | | | | | | | | | | | | | **-21.5** | **1.5** | ORAU; Damon *et al.* 1995; Menjo *et al.* 2010 |
| 1010 | 4 | -16.2 | 3.8 | -15.2 | 3.3 | -17.0 | 2.0 | | | | | | | | | | | | | | | **-16.5** | **1.6** | Damon *et al.* 1995; Menjo *et al.* 2011 |
| 1011 | 5 | | | -10.0 | 3.5 | -19.7 | 2.0 | | | | | | | | | | | | | | | **-17.3** | **1.7** | Damon *et al.* 1995; Menjo *et al.* 2012 |
| 1012 | 6 | | | -17.1 | 3.3 | -16.5 | 2.0 | | | | | | | | | | | | | | | **-16.7** | **1.7** | Damon *et al.* 1995; Menjo *et al.* 2013 |
| 1013 | 7 | | | -11.1 | 3.2 | -15.3 | 2.0 | | | | | | | | | | | | | | | **-14.1** | **1.7** | Damon *et al.* 1995; Menjo *et al.* 2014 |
| 1014 | 8 | | | -16.7 | 3.0 | -18.5 | 2.8 | | | | | | | | | | | | | | | **-17.7** | **2.1** | Damon *et al.* 1995; Menjo *et al.* 2015 |
| 1015 | 9 | | | -11.8 | 4.1 | -12.1 | 2.8 | | | | | | | | | | | | | | | **-12.0** | **2.3** | Damon *et al.* 1995; Menjo *et al.* 2016 |
| 1016 | 10 | | | -18.3 | 3.4 | -17.0 | 2.8 | | | | | | | | | | | | | | | **-17.5** | **2.2** | Damon *et al.* 1995; Menjo *et al.* 2017 |
| 1050 | -4 | | | | | -11.5 | 2.8 | | | | | | | | | | | | | | | **-11.5** | **2.8** | Menjo *et al.* 2005 |
| 1052 | -2 | | | | | -7.8 | 2.8 | | | | | | | | | | | | | | | **-7.8** | **2.8** | Menjo *et al.* 2005 |
| 1054 | 0 | | | | | -2.2 | 2.8 | | | | | | | | | | | | | | | **-2.2** | **2.8** | Menjo *et al.* 2005 |
| 1056 | 2 | | | | | -7.5 | 2.8 | | | | | | | | | | | | | | | **-7.5** | **2.8** | Menjo *et al.* 2005 |
| 1058 | 4 | | | | | -8.4 | 2.8 | | | | | | | | | | | | | | | **-8.4** | **2.8** | Menjo *et al.* 2005 |
| 1060 | 6 | | | | | -7.1 | 2.8 | | | | | | | | | | | | | | | **-7.1** | **2.8** | Menjo *et al.* 2005 |
| 1062 | 8 | | | | | -10.1 | 2.8 | | | | | | | | | | | | | | | **-10.1** | **2.8** | Menjo *et al.* 2005 |
| 1064 | 10 | | | | | -13.3 | 2.8 | | | | | | | | | | | | | | | **-13.3** | **2.8** | Menjo *et al.* 2005 |
| 1567 | -5 | | | | | | | | | 4.2 | 1.7 | | | | | | | | | | | **4.2** | **1.7** | Reimer *et al.* 2013 |
| 1568 | -4 | | | | | | | | | 4.4 | 2.5 | | | | | | | | | | | **4.4** | **2.5** | Reimer *et al.* 2013 |
| 1569 | -3 | | | | | | | | | 5.6 | 2.3 | | | | | | | | | | | **5.6** | **2.3** | Reimer *et al.* 2013 |
| 1570 | -2 | | | | | | | | | 3.5 | 1.7 | | | | | | | | | | | **3.5** | **1.7** | Reimer *et al.* 2013 |
| 1571 | -1 | | | | | | | | | 5.4 | 2.5 | | | | | | | | | | | **5.4** | **2.5** | Reimer *et al.* 2013 |
| 1572 | 0 | | | | | | | | | 6.3 | 2.3 | | | | | | | | | | | **6.3** | **2.3** | Reimer *et al.* 2013 |
| 1573 | 1 | | | | | | | | | 7.9 | 2.5 | | | | | | | | | | | **7.9** | **2.5** | Reimer *et al.* 2013 |
| 1574 | 2 | | | | | | | | | 4.0 | 1.2 | | | | | | | | | | | **4.0** | **1.2** | Reimer *et al.* 2013 |
| 1575 | 3 | | | | | | | | | 3.9 | 2.3 | | | | | | | | | | | **3.9** | **2.3** | Reimer *et al.* 2013 |
| 1576 | 4 | | | | | | | | | 1.8 | 2.2 | | | | | | | | | | | **1.8** | **2.2** | Reimer *et al.* 2013 |
| 1577 | 5 | | | | | | | | | 3.3 | 2.5 | | | | | | | | | | | **3.3** | **2.5** | Reimer *et al.* 2013 |
| 1578 | 6 | | | | | | | | | 5.2 | 2.3 | | | | | | | | | | | **5.2** | **2.3** | Reimer *et al.* 2013 |
| 1579 | 7 | | | | | | | | | 2.4 | 2.2 | | | | | | | | | | | **2.4** | **2.2** | Reimer *et al.* 2013 |
| 1580 | 8 | | | | | | | | | 3.6 | 2.3 | | | | | | | | | | | **3.6** | **2.3** | Reimer *et al.* 2013 |
| 1581 | 9 | | | | | | | | | -0.3 | 2.2 | | | | | | | | | | | **-0.3** | **2.2** | Reimer *et al.* 2013 |
| 1582 | 10 | | | | | | | | | 3.1 | 2.3 | | | | | | | | | | | **3.1** | **2.3** | Reimer *et al.* 2013 |
| 1599 | -5 | | | | | | | | | -1.0 | 2.2 | | | | | | | | | | | **-1.0** | **2.2** | Reimer *et al.* 2013 |
| 1600 | -4 | | | | | | | | | 0.8 | 1.6 | | | | | | | | | | | **0.8** | **1.6** | Reimer *et al.* 2013 |
| 1601 | -3 | | | | | | | | | -1.4 | 1.7 | | | | | | | | | | | **-1.4** | **1.7** | Reimer *et al.* 2013 |
| 1602 | -2 | | | | | | | | | -2.0 | 2.5 | | | | | | | | | | | **-2.0** | **2.5** | Reimer *et al.* 2013 |
| 1603 | -1 | | | | | | | | | -4.3 | 2.5 | | | | | | | | | | | **-4.3** | **2.5** | Reimer *et al.* 2013 |
| 1604 | 0 | | | | | | | | | -3.2 | 1.1 | | | | | | | | | | | **-3.2** | **1.1** | Reimer *et al.* 2013 |
| 1605 | 1 | | | | | | | | | -5.4 | 1.8 | | | | | | | | | | | **-5.4** | **1.8** | Reimer *et al.* 2013 |
| 1606 | 2 | | | | | | | | | -4.7 | 1.5 | | | | | | | | | | | **-4.7** | **1.5** | Reimer *et al.* 2013 |
| 1607 | 3 | | | | | | | | | -4.8 | 2.6 | | | | | | | | | | | **-4.8** | **2.6** | Reimer *et al.* 2013 |
| 1608 | 4 | | | | | | | | | -3.2 | 1.8 | | | | | | | | | | | **-3.2** | **1.8** | Reimer *et al.* 2013 |
| 1609 | 5 | | | | | | | | | -3.4 | 1.7 | | | | | | | | | | | **-3.4** | **1.7** | Reimer *et al.* 2013 |
| 1610 | 6 | | | | | | | | | -2.7 | 2.5 | | | | | | | | | | | **-2.7** | **2.5** | Reimer *et al.* 2013 |
| 1611 | 7 | | | | | | | | | -6.4 | 1.8 | | | | | | | | | | | **-6.4** | **1.8** | Reimer *et al.* 2013 |
| 1612 | 8 | | | | | | | | | -1.4 | 2.5 | | | | | | | | | | | **-1.4** | **2.5** | Reimer *et al.* 2013 |
| 1613 | 9 | | | | | | | | | -5.2 | 2.2 | | | | | | | | | | | **-5.2** | **2.2** | Reimer *et al.* 2013 |
| 1614 | 10 | | | | | | | | | -2.3 | 1.2 | | | | | | | | | | | **-2.3** | **1.2** | Reimer *et al.* 2013 |

| Name | Lab Ref | Calendar Year | $^{14}C$ Date (BP) | ± (σ) | $\delta^{13}C$ (PDB) |
|---|---|---|---|---|---|
| SB | OxA-30923 | 6 BCE | 2007 | 29 | -23.7 |
| | OxA-30924 | 5 BCE | 2021 | 31 | -24.7 |
| | OxA-30925 | 4 BCE | 2019 | 30 | -25.0 |
| | OxA-30926 | 3 BCE | 2024 | 29 | -24.3 |
| | OxA-30927 | 2 BCE | 1982 | 30 | -22.5 |
| | OxA-31890 | 2 BCE | 2044 | 26 | -22.5 |
| | OxA-30928 | 1 BCE | 2025 | 29 | -24.0 |
| SN185 | OxA-30877 | 183 CE | 1876 | 29 | -20.4 |
| | OxA-30878 | 184 CE | 1970 | 29 | -20.2 |
| | OxA-30879 | 185 CE | 1898 | 28 | -19.8 |
| | OxA-30880 | 186 CE | 1981 | 28 | -19.8 |
| | OxA-30881 | 187 CE | 1881 | 29 | -19.9 |
| | OxA-30882 | 188 CE | 1969 | 29 | -19.5 |
| SN1006 | OxA-30888 | 1004 CE | 1026 | 28 | -25.3 |
| | OxA-30889 | 1005 CE | 1110 | 28 | -24.9 |
| | OxA-30559 | 1006 CE | 1018 | 26 | -24.7 |
| | OxA-30964 | 1007 CE | 1058 | 31 | -24.7 |
| | OxA-30965 | 1008 CE | 1107 | 28 | -24.4 |
| | OxA-30966 | 1009 CE | 1114 | 29 | -24.6 |

**Table S2. Further details of the radiocarbon dates obtained by Oxford for this study.**